\documentclass{article}
\font\cero=cmss10 scaled 1728 
\usepackage{tensor}
\usepackage{verbatim}
\usepackage{amsfonts}
\usepackage{graphicx}
\usepackage{subcaption}
\DeclareCaptionLabelFormat{figlabel}{Fig.#2}
\usepackage{hyperref}

\usepackage{amssymb}
\usepackage{float}
\usepackage{amsmath}
\usepackage[dvipsnames]{xcolor}
\begin{document}
\begin{flushleft}
{\cero Hypercomplex formulation of dissipative scalar electrodynamics and phase transitions}\\
\end{flushleft}
{\sf B.R. López-Raymundo} and \noindent {\sf R. Cartas-Fuentevilla} \\ 
\noindent {\it Instituto de F\'{\i}sica, Universidad Aut\'onoma de Puebla,
Apartado postal J-48 72570 Puebla Pue., M\'exico}; 

E-mail: blopezr@ifuap.buap.mx, rcartas@ifuap.buap.mx

ABSTRACT: A hypercomplex formulation of scalar electrodynamics is developed, in which dissipation emerges as an intrinsic dynamic property dictated by the extended local gauge symmetry $U(1) \times SO(1,1)$. By constructing an effective free-energy functional, we derive the generalized Ginzburg-Landau equations and analyze the thermodynamic stability of the system. Our analysis of the vacuum state up to the fourth order reveals that the constraint on the coefficients associated with the invariants prohibits the coexistence of mixed states, formally establishing the existence of a bicritical point. Consequently, upon crossing the critical temperature, the system is forced to undergo a discontinuous first-order phase transition. Furthermore, we demonstrate that the hyperbolic symmetry induces a structural instability when the effective quartic couplings take negative values, rendering the free energy unbounded from below and creating asymptotic directions of instability.    \\

\noindent KEYWORDS: Dissipative scalar electrodynamics;  Ginzburg-Landau; phase transition; hypercomplex ring. 

\section {Introduction}
\label{intro}
Scalar Quantum Electrodynamics (SQED) constitutes one of the methodological pillars of quantum field theory. Historically, it served as the fundamental conceptual framework under which the formulation of spinor electrodynamics was developed, demonstrating that the requirement of local phase invariance based on the $U(1)$ symmetry group naturally generates electromagnetic interactions \cite{dirac,weyl}. At a macroscopic level, this abelian formulation transcends elementary particle physics and finds profound applications in multiple branches of physics. In condensed matter, for instance, the successful Ginzburg-Landau phenomenological theory for superconductivity finds its rigorous microscopic justification precisely in the structure of a scalar electrodynamics with a spontaneously broken $U(1)$ symmetry \cite{landau1,ginzburg}. Furthermore, recent research has explored field theories based on the hypercomplex ring formulation; this approach introduces a novel hyperbolic algebraic unit $j$ which, in conjunction with the standard complex unit $i$ and their product, endows the formulation with a hypercomplex structure. Within this framework, several works have incorporated hyperbolic numbers into relativistic quantum mechanics, as well as in the representation of the Schrödinger and Klein-Gordon equations \cite{UR,hyperbolicS,U4}. Notably, in \cite{U1} it is demonstrated that a gravitoelectromagnetic gauge symmetry emerges by introducing a hyperbolic unit into the complex phase symmetry of electromagnetism, an analytical tool that has also been employed to study the hermiticity of the Poincaré mass operator \cite{U3}. Specifically, the extension offered by the hypercomplex formulation allows for a natural description of dissipative quantum field theory, as proposed in previous works \cite{ssbh,Hyper1,Hyper2,Hyper3}. Under this formalism, dissipation ceases to be an ad hoc phenomenological term added to the equations of motion; instead, it is interpreted as an intrinsic dynamic property of the system. This is achieved by redefining the field in terms of hypercomplex quantities, a process mathematically analogous to the doubling of degrees of freedom, wich is characteristic in the description of open systems. Consequently, the Lagrangian governing the dissipative system is constructed as an invariant quantity under the $U(1) \times SO(1, 1)$ group, operating under a combination of standard complex and hyperbolic rotations that guarantee a coherent relativistic description of the interaction between the physical system and its reservoir; the rigorous treatment of dissipative dynamics is essential in modern physics. In the realm of open quantum systems and quantum information, these theories are fundamental to describe environment-induced decoherence and the irreversible evolution of macroscopic quantum states \cite{breuer,caldeira}. In the high-energy regime, the formulation of relativistic dissipative hydrodynamics is indispensable to understand thermalization and viscosity effects in the evolution of the quark-gluon plasma generated in heavy-ion collisions \cite{romats,israel}. Likewise, at cosmological scales, dissipative effects in the dynamics of scalar fields provide natural mechanisms to model particle production, bulk viscosity in the expansion of the universe, and the dynamics of inflation \cite{bastero,bartrum}. Motivated by this background, and continuing this line of research, the first sections of this work establish the formulation of a hypercomplex scalar electrodynamics. By requiring the theory to preserve invariance under the extended local symmetry $U(1) \times SO(1,1)$, dissipation emerges as an intrinsic and relativistically consistent dynamics, strictly dictated by the gauge coupling of hyperbolic rotations. This structure serves as the fundamental microscopic basis for developing a thermodynamic theory of phase transitions applicable to dissipative systems. Analogous to how standard SQED underpins the Landau theory for superconductors, this article constructs an effective hypercomplex free-energy functional designed to map the total state of the coupled system. From this functional, the application of the strict variational principle leads to the analytical derivation of a hypercomplex version of the Ginzburg-Landau equations. These generalized expressions carry profound physical implications, as they primarily dictate the penetration profiles of electromagnetic interactions and the distribution of macrocurrents associated with the thermodynamic equilibrium of an intrinsically dissipative system.

Once the global dynamic conditions are formalized, the analytical focus of this study concentrates on the vacuum manifold; uder these conditions, the parameter $\gamma$, initially introduced with an algebraic motivation to reduce the degrees of freedom of the hypercomplex field, is directly promoted to a macroscopic thermal parameter. This extension reveals how interactions impose a strict directional competition in the condensation of the fields; specifically, the thermodynamic analysis, based on the expansion of the order parameter up to the fourth order, demonstrates that the geometric constraint of the invariants imposes a strictly negative quartic discriminant. This singularity prohibits any mixed-state configuration, thereby establishing the formal existence of a bicritical point and dictating the occurrence of first-order phase transitions.

In the next section formally describes the notion of a hypercomplex ring, detailing the mathematical elements and properties that constitute the foundation of our model. Based on this, Section 3 addresses the construction of the dissipative system by defining a hypercomplex field that unifies a charged subsystem and its reservoir; this combined dynamic system is reduced to exactly two degrees of freedom via a parametric transformation. In Section 4, a hypercomplex formulation of scalar electrodynamics is developed by requiring the theory to preserve the extended local gauge symmetry $U(1) \times SO(1,1)$. Finally, Section 5 integrates a macroscopic analysis through the phenomenological Landau theory of phase transitions, which is based on constructing a free-energy functional strictly from the invariants of the symmetry group. Within this framework, Section 5.1 applies the equilibrium conditions to the functional to derive the generalized Ginzburg-Landau equations for the dissipative system. In particular, Section 5.2 examines the nature of phase transitions in the vacuum state and the fundamental role of the thermal parameter $\gamma$, primarily determining the scenarios of structural instability arising from the thermal dependence of the fourth-order coefficients.
\section {Incorporating the hypercomplex ring}
\label{hyrot}
The ordinary complex numbers can be extended to the hypercomplex numbers $z$ and it constitutes a conmutative ring $\mathbb{H}$ \cite{U1};
\begin{eqnarray}
     z =  x + iy +jv + ijw, \quad \overline{z}  =  x - iy -jv + ijw, \quad  x,y,v,w \in \mathbb{R} 
          \label{ring}
\end{eqnarray} 
with the properties of the hyperbolic unit: $j^{2}=1$, $\bar j=-j$ and the usual properties for the standard complex unit $i^{2}=-1$ and $\bar i=-i$. In addition, the hybrid number $ij$ satisfies $(ij)^2 = -1$ and $\overline{ij} = ij$. The hyperbolic numbers $\mathbb{P}$ and the complex numbers $\mathbb{C}$ are subsets of the ring $\mathbb{H}$. Therefore, the square modulus of a hypercomplex number is defined as follows:
\begin{equation}
     z \overline{z} = x^{2} + y^{2} -v^{2} - w^{2} +2ij (xw-yv),
     \label{square}
\end{equation} 
this quantity is not a real number; instead it is a Hermitian number. Furthermore, it is invariant under the usual circular rotation $e^{i\theta}$ represented by the group $U(1)$ and under hyperbolic rotations $e^{j\chi} = cosh\chi + jsinh\chi$,  where $\chi$ is a non-compact real number and which can be represented by the group  $SO(1,1)$. Therefore, the modulus is invariant under the full phase $e^{j\theta}e^{j\chi}$, which corresponds to the group $U(1)\times SO(1,1)$. \\
The definition of hypercomplex number can be rewritten by restricting ourselves to two degrees of freedom, that is,  can be defined in terms of two real quantities by means of the appropriate identification: $x=\gamma w$ and $y=\gamma v$, with $\gamma$ as a real parameter. In this way, the equations \eqref{ring} and \eqref{square}  is reduced to \cite{ssbh}:
\begin{equation}
     z= (\gamma +ij)w+(i\gamma +j)v , \quad \overline{z} = (\gamma +ij)w-(i\gamma +j)v,\quad  z\overline{z} = (\gamma^{2}-1)(v^{2}+w^{2}) + 2ij\gamma (w^{2}-v^{2});
     \label{cartesian}
\end{equation}
thus the equation is invariant under the exchange of the variables $v\leftrightarrow  w $ and under the change  $\gamma \to -\gamma$. 
Additionally, it is also possible to express the number $z$ in a general polar form as
\begin{equation}
    z=\rho e^{i\eta}e^{j\xi}, \qquad \rho=\rho_R+ij\rho_H;
    \label{polar}
\end{equation}
with $\eta\in (0,2\pi]$, $\eta\in\mathbb{R}$ and $\rho$ as a Hermitian quantity with $\rho_R$ and $\rho_H$ $\in\mathbb{R}$. From the Cartesian form \eqref{cartesian} and polar form \eqref{polar}, we can find the following relationships:
\begin{equation}
    \rho_R^2-\rho_H^2=(\gamma^2-1)(w^2+v^2), \qquad \rho_H\rho_R=\gamma(w^2-v^2);
\end{equation}
from which, by solving, we can find a direct expression of the quantities $\rho_R$ and $\rho_H$ in terms of $w$, $v$, and $\gamma$:
\begin{eqnarray}
    \rho_R^2 \!\! & = & \!\! \frac{(\gamma^2 - 1)(w^2 + v^2) + \sqrt{\left[ (\gamma^2 - 1)(w^2 + v^2) \right]^2 + 4\gamma^2(w^2 - v^2)^2}}{2}, \nonumber \\ \rho_H^2\!\! & =& \!\!  \frac{(1-\gamma^2)(w^2 + v^2) + \sqrt{\left[ (\gamma^2 - 1)(w^2 + v^2) \right]^2 + 4\gamma^2(w^2 - v^2)^2}}{2}, 
    \end{eqnarray}
and similarly, by equating \eqref{cartesian} and \eqref{polar} and after a series of steps, we can obtain the expressions for the phases $\eta$ and $\xi$ given by the following formulas:
\begin{eqnarray}
    \eta \!\! & = & \!\! \frac{1}{2} \tan^{-1} \left[ \frac{2vw(\gamma^2 - 1)}{(\gamma^2 + 1)(w^2 - v^2)} \right],  \\ \nonumber \\ \xi \!\! & =& \!\! \frac{1}{4} \ln \left[ \frac{(\gamma^2 + 1)(w^2 + v^2) + 4\gamma vw}{(\gamma^2 + 1)(w^2 + v^2) - 4\gamma vw} \right].
\end{eqnarray}
Now, applying hyperbolic-circular rotation to the hypercomplex number \(z\) in the polar form \eqref{polar}, we get:
\begin{eqnarray}
e^{i\theta} e^{j\chi} z \!\! & = & \!\!\rho e^{i(\theta + \eta)} e^{j(\chi + \xi)}\nonumber \\ \!\! & =& \!\! \big[ \rho_R (\cos\theta\cos\eta - \sin\theta\sin\eta) (\cosh\chi\cosh\xi + \sinh\chi\sinh\xi)  \nonumber \\ \!\! & &\!\!-\rho_H(\sin\theta\cos\eta + \cos\theta\sin\eta) (\sinh\chi\cosh\xi + \cosh\chi\sinh\xi)\big]\nonumber \\ \!\! & &\!\!+i \big[ \rho_R (\sin\theta\cos\eta + \cos\theta\sin\eta) (\cosh\chi\cosh\xi + \sinh\chi\sinh\xi) \nonumber \\ \!\! & &\!\!  +\rho_H (\cos\theta\cos\eta - \sin\theta\sin\eta) (\sinh\chi\cosh\xi + \cosh\chi\sinh\xi) \big]\nonumber \\ \!\! & &\!\!+j \big[ \rho_R  (\cos\theta\cos\eta - \sin\theta\sin\eta) (\sinh\chi\cosh\xi + \cosh\chi\sinh\xi) \nonumber \\ \!\! & &\!\!  - \rho_H  (\sin\theta\cos\eta + \cos\theta\sin\eta) (\cosh\chi\cosh\xi + \sinh\chi\sinh\xi) \big]\nonumber \\ \!\! & &\!\!+ij \big[\rho_R (\sin\theta\cos\eta + \cos\theta\sin\eta) (\sinh\chi\cosh\xi + \cosh\chi\sinh\xi) \nonumber \\ \!\! & &\!\! + \rho_H (\cos\theta\cos\eta - \sin\theta\sin\eta) (\cosh\chi\cosh\xi + \sinh\chi\sinh\xi) \big].
\label{rotation}
\end{eqnarray}
From the definition \eqref{cartesian}  we have that if \(\gamma = 0\), the number $z$ reduces to an ordinary number $z = j(v + i w)$, but if \(\gamma \neq 0\), the norm $z\bar{z}$ will necessarily contain a hybrid term $ij$, so it will not be a purely real quantity. On the other hand, by choosing \(\gamma^2 = 1\), the norm will be purely a hybrid quantity $ij$, that is, a pure hypercomplex number. However, if \(\gamma^2 \neq 1\), this represents the non-trivial combination of rotations \eqref{rotation}, and such a deformation will have profound effects on the dynamics of the field theory construted. Some other properties can be consulted in \cite{ssbh,sobczyk}.

\noindent It is pertinent to emphasize that, while the introduction of the parameter $\gamma$ is originally for reducing the components of the hypercomplex variable, its relevance will manifest in the following sections by endowing it with a fundamental physical meaning. The emergence of macroscopic thermodynamic properties from purely geometric structures constitutes one of the deepest conceptual foundations of the theory. Historically, various rigorous formalisms have demonstrated that temperature does not need to be introduced as an external phenomenological variable, but can be directly encoded in the underlying symmetries. A notable example is the imaginary time formalism, developed to directly connect field theory with statistical mechanics, where temperature arises as a strict consequence of the compactification of Euclidean spacetime \cite{matsubara}.
Analogously, in the Unruh effect, the temperature is analytically deduced from the geometry of spacetime considering that the vacuum state of a quantum field theory on Minkowski space looks like a thermal equilibrium state for a uniformly accelerated observer \cite{crispino,martinetti}. In Thermal Field Dynamics (TFD), the macroscopic thermal vacuum is defined through Bogoliubov transformations that connect the degrees of freedom of the physical system with those of a dual space (the thermal bath); these operations mathematically correspond to hyperbolic rotations for bosonic system and circular ones for fermionic systems \cite{khanna}. Inspired by these precedents, in our model, the parameter $\gamma$, that emerges from the geometric constraint based on circular and hyperbolic rotations of the symmetry $U(1)\times SO(1,1)$ to describe charged bosonic systems, can be naturally identified with temperature. This establishes a rigorous connection between the algebraic structure governing microscopic dissipative dynamics and the macroscopic framework of Landau phase transitions.
\section{ Dissipative Lagrangian in the hypercomplex formulation } 
Diverse works describing dissipative quantum systems and thermal fields have been based on doubling the degrees of freedom \cite{QD, Galley, Saha}; previous studies have adapted this idea to the hypercomplex formalism \cite{Hyper1, Hyper2}. The basis of this work is to take up these ideas and consider a physical system composed of a real charged field $\Psi$ as the system of interest and a real charged field $\Phi$ as the thermal bath or environment. The system as a set will be represented by a hypercomplex field $\Omega$, wich can be written in the form:
\begin{equation}
\Omega=\Psi+j\Phi=\psi_1+i\psi_2+j\phi_1+ij\phi_2,
\label{Hyperfield}
\end{equation}
with the standard charged fields,
\begin{equation}
    \Psi=\psi_1+i\psi_2,\qquad  \Phi=\phi_1+i\phi_2; \qquad \quad \psi_1,\psi_2,\phi_1,\phi_2\in \mathbb{R}.
\end{equation}
Based on this definition, we can construct the following dissipative global $U(1) \times SO(1,1)$ invariant Lagrangian on a d-dimensional Minkowskian background:
\begin{equation}
\mathcal{L}=\partial_\mu\Omega\partial^\mu\overline\Omega - V(\Omega\overline\Omega),
    \label{Lh1}
\end{equation}
in which the potential term $V(\Omega\overline\Omega)$ is also included,
\begin{equation}
V(\Omega\overline\Omega)=\frac{1}{2}m^{2}\Omega\overline{\Omega}+\frac{\lambda}{4!}(\Omega\overline{\Omega})^2,
    \label{V1}
\end{equation}
where the mass term has the general form  $m^2=m_R^2+ijm_H^2$ and similarly for the self-interaction term $\lambda=\lambda_R+ij\lambda_H$, with ($m_R^2,m_H^2,\lambda_R,\lambda_H$) real parameters \cite{ssbh, Hyper2}. The subscripts $R$ and $H$ denote the contributions from the system of interest and the thermal bath, respectively. Note also that this potential is an invariant of the group $U(1) \times SO(1,1)$ according to the principles discussed in the previous section. While the kinetic term of the \eqref{Lh1}  take the form
\begin{equation}
    \partial_\mu\Omega \partial^\mu\overline\Omega=(\partial_\mu\psi_1)^2+(\partial_\mu\psi_2)^2-(\partial_\mu\phi_1)^2-(\partial_\mu\phi_2)^2+2ij(\partial_\mu\phi_1\partial_\mu\psi_2-\partial_\mu\phi_2\partial_\mu\psi_1),
    \label{kinetic}
\end{equation}
in which the term that contributes to the interaction between the two charged fields is contained in the hybrid part ($ij$). 
The analytical development of this work on the dissipative model begins with a fundamental constraint between the fields $\Psi$ and $\Phi$ in \eqref{Hyperfield}, reducing the real dynamical variables using the following transformation:
\begin{equation}
    \Psi=\gamma e^{i\pi/2}\Phi^*,
    \label{trans1}
\end{equation}
this leads directly to the reduced hypercomplex form given in equation \eqref{cartesian}, now with:
\begin{equation}
    \Omega\overline{\Omega}= (\gamma^2 - 1)(\phi_2^2 + \phi_1^2)
  + 2 i j \gamma (\phi_2^2 - \phi_1^2),
  \label{norm1}
\end{equation}
where we have eliminated the pair $(\psi_1,\psi_2)$ to favor of the pair $(\phi_1,\phi_2)$.
The transformation \eqref{trans1} establishes a strict relationship in which the field amplitudes are parametrically coupled via $\gamma$, which leads to bounding the potencial \eqref{V1} from below and will also play an important role as the thermal variable of the theory. Furthermore, the phase term under a $\pi/2$ and complex conjugation ensure a detailed balance of the vacuum energy. Based on these considerations, we establish the basic thermodynamic framework, focusing strictly on the steady-state limit, which is invariant under global transformations of the rotation symmetry groups $U(1) \times SO(1,1)$ of the form $\Omega\to\Omega e^{i\theta}e^{j\chi}$. 
\\
In this framework, it can also be shown that by simply considering the $U(1)\times SO(1,1)$ invariant potential \eqref{V1}, dissipative effects emerge naturally through the extension of the general solution for the equations of motion in the complex-hyperbolic plane. Specifically, by promoting the pair ($\omega,\vec{k}$) to Hermitian quantities, dissipation arises by structuring an ansatz with hypercomplex phases of the form:
\begin{equation}
 \Omega = e^{i [ (w_1 + ij w_2)t - (\vec k_1 + ij \vec k_2) \cdot \vec x ]}, \qquad \omega\to \omega_1+ij\omega_2, \quad \vec{k}\to \vec{k}_1+ij\vec{k}_2;  
\end{equation}
this solution allows the D'Alembert operator to remain unchanged as an invariant under hyperbolic rotations where energy-momentum dissipation exists between the subsystem and the reservoir. Such results can be found in  \cite{Hyper3}, where it is explored how these extended solutions modify the dispersion relations.
As we shall see in this work, in this interacting scenario thermodynamic equilibrium is not defined by a trivial asymptotic isolation, but rather by a steady-state of detailed balance. Far from heuristic interpretations, this steady-state balance is formalized by prioritizing the extremization of the functionals. This strict variational principle will be developed in the following sections, where we will proceed to derive the system of equations for the minima of the coupled system, further incorporating electromagnetic interactions. 
\section { Hypercomplex scalar electrodynamics}  
\label{hsed}
The development of quantum electrodynamics and its scalar counterpart were consolidated by requiring that the global phase symmetry to be postulated as a local symmetry. Following these same principles for gauge symmetry, we proceed to construct a scalar electrodynamics in a hypercomplex version.
After establishing the conditions in a stationary limit, the next formal step consists of endowing the dissipative model with dynamic interactions consistent with the principles of relativity and field theory. In our formalism, the scalar doublet is subject to an internal symmetry governed by the group $U(1) \times SO(1,1)$. We will therefore require that the physics of the system remain unchanged if the phase combinations vary independently at each point in spacetime. This translates into subjecting the hypercomplex field to the following local transformation: 
\begin{equation}
\Omega(x) \longrightarrow e^{i\theta(x)} e^{j\chi(x)} \Omega(x);
\qquad
\bar{\Omega}(x) \longrightarrow e^{-i\theta(x)} e^{-j\chi(x)} \bar{\Omega}(x).
\label{transhiper}
\end{equation}
With this local transformation, the direct application of the ordinary derivative to the kinetic term yields derivatives with respect to the parameters $(\theta,\chi)$, and as a result, these additional terms break the invariance of the Lagrangian \eqref{Lh1}. To restore the fundamental symmetry and preserve the structure of the theory, it is imperative to replace the ordinary derivative with a covariant derivative operator $D_\mu$, which is proposed as follows:
\begin{equation}
     D_\mu \Omega =\left[ 
  \partial_\mu+ ie \mathcal{C}_\mu  
\right]\Omega=\left[ 
  \partial_\mu+ ie( \mathcal{A}_\mu-ij\mathcal{B}_\mu)  
\right]\Omega, \qquad \mathcal{C}_\mu=\mathcal{A}_\mu-ij\mathcal{B}_\mu;
\label{CovariantD}
\end{equation}
where $\mathcal{C}_{\mu}$ is the so-called connection and $e$ is the coupling constant. The expression \eqref{CovariantD} shows that the covariant derivative incorporates two independent gauge contributions, each associated with one of the generators of the hypercomplex algebra. The quantities $\mathcal{A}_\mu$ and $\mathcal{B}_\mu$ can therefore be interpreted as the gauge fields associated with the subsystem and the environment, respectively and the transformation rules obtained are as follows:
\begin{equation}
    \begin{split}
        \mathcal{A}'_\mu=\mathcal{A}_\mu-\frac{1}{e} \partial_\mu \theta, \qquad  \mathcal{B}'_\mu=\mathcal{B}_\mu-\frac{1}{e}\partial_\mu \chi.
    \label{A2}
    \end{split}
\end{equation}
As it has been well established in the literature \cite{peskin}, the dynamics of these mediating fields can be deduced geometrically by evaluating the commutator of the covariant derivatives, thereby revealing the field intensity tensor associated with this extended local symmetry:
\begin{equation}
    \begin{split}
        [D_\mu, D_\nu] \, \Omega &=ie(F_{\mu\nu}-ijG_{\mu\nu})\Omega, \quad \quad F_{\mu\nu} = \partial_\mu \mathcal{A}_\nu-\partial_\nu \mathcal{A}_\mu, \quad G_{\mu\nu} = \partial_\mu \mathcal{B}_\nu-\partial_\nu \mathcal{B}_\mu;
    \end{split}
 \end{equation}  
where $H_{\mu \nu}= F_{\mu\nu}-ijG_{\mu\nu}$. Since the symmetry group is strictly Abelian, the commutators of the generators cancel out, so the stress tensors $F_{\mu\nu}$ and $G_{\mu\nu}$ are associated with the subsystem and reservoir sectors respectively.
If we consider gauge fields to be physical fields, we must add the kinetic energy term to the total Lagrangian; this must be a quadratic invariant constructed as $-\frac{1}{4} H_{\mu\nu}H^{\mu\nu}$. By incorporating interactions mediated by gauge fields, we obtain a formulation that strictly preserves the local $U(1) \times SO(1,1)$ symmetry and establishes the fundamental framework for describing the dissipative dynamics of the fields, thus the total Lagrangian density of the system is constructed as 
\begin{eqnarray}
\mathcal{L}_D(\Omega,\overline{\Omega},\mathcal{C}_\mu) \!\! & = & \!\!D_\mu \Omega D^\mu\overline{\Omega} -\frac{1}{4} H_{\mu\nu}H^{\mu\nu}-V( \Omega\overline{\Omega}) \nonumber \\ \!\!  & = & \!\!\left[ \partial_\mu+ ie(\mathcal{A}_\mu-ij\mathcal{B}_\mu) \right]\Omega \left[ 
  \partial^\mu- ie(\mathcal{A}^\mu-ij\mathcal{B}^\mu)  \right] \overline{\Omega}  \nonumber \\ \!\! & & \!\! -\frac{1}{4}(F_{\mu\nu} -ijG_{\mu\nu}) (F^{\mu\nu} -ijG^{\mu\nu}) \nonumber \\ \!\! & & \!\!-\frac{1}{2}(m^2_R+ijm^2_H)\Omega\overline{\Omega}-\frac{1}{4!} (\lambda_R+ij\lambda_H)(\Omega\overline{\Omega})^2,
  \label{LD}
\end{eqnarray}
and the corresponding equations of motion are,
\begin{equation}
    \left[ \partial_\mu+ ie(\mathcal{A}_\mu-ij\mathcal{B}_\mu) \right] \left[ 
  \partial^\mu+ ie(\mathcal{A}^\mu-ij\mathcal{B}^\mu)  \right] \Omega+\frac{1}{12}\left[(\lambda_R+ij\lambda_H)\Omega\overline{\Omega} + 6(m^2_R+ijm^2_H) \right] \Omega=0,
\end{equation}
\begin{equation}
    \partial_\nu \left[(\partial^\mu \mathcal{A}^\nu-\partial^\nu \mathcal{A}^\mu)-ij
    (\partial^\mu \mathcal{B}^\nu-\partial^\nu \mathcal{B}^\mu) \right]=ie \left[ \Omega  
  (\partial^\mu- ie(\mathcal{A}^\mu-ij\mathcal{B}^\mu))  \overline{\Omega}- \overline{\Omega} (\partial^\mu+ ie(\mathcal{A}^\mu-ij\mathcal{B}^\mu)) \Omega \right].
\end{equation}
Once the Lagrangian structure of scalar electrodynamics with local interactions has been established, the focus now turns to analyzing the macroscopic behavior of the system in the presence of possible thermal variations. To this end, it is necessary to formulate a phenomenological theory of phase transitions based on Landau’s original postulates, the essence of which lies in relating these changes of state to the spontaneous symmetry breaking of the system \cite{Landau1936}. Within this framework, explicitly structured on the algebraic basis of the hypercomplex ring, the model describes the transition from a disordered, highly symmetric configuration to an ordered state as a consequence of the breaking of the local group $U(1) \times SO(1,1)$.
\section{A hypercomplex formulation of Landau theory of phase transitions}
The standard methodology for implementing these principles involves introducing a Ginzburg-Landau functional. Derived through a mean-field approach that integrates out microscopic degrees of freedom to describe the system via a continuous macroscopic field, and this functional acts as a generalized effective free energy. It allows us to characterize the accessible thermodynamic states of the model, where its minimization from first principles rigorously dictates the equilibrium configurations. A fundamental requirement for its construction is strict invariance under all underlying group transformations \cite{Landau1958,olmsted,Kardar}.
From this, it follows directly that all possible invariants of the theory correspond to the even powers of the quantity given by \eqref{V1}, the form of the covariant derivative \eqref{CovariantD}, and the kinetic terms of the electromagnetic interaction given by the contraction of the field intensity tensor. Consequently, by selecting the hypercomplex scalar doublet $\Omega$ as the macroscopic-order parameter and from \eqref{LD}, the general hypercomplex functional in d-dimensions is given by:
\begin{eqnarray}
   F[\Omega, \overline{\Omega}, \vec{A}_1, \vec{A_2}] \!\! & = & \!\! \int d^d x \mathcal{F}_{GL} \nonumber \\ \!\! & = & \!\!  \int d^d x \Biggl\{ e^2 (U_1 - ij U_2)^2 \Omega \overline{\Omega} + [\nabla - ie(\vec{A}_1 - ij \vec{A}_2)] \Omega \cdot [\nabla + ie(\vec{A}_1 - ij \vec{A}_2)] \overline{\Omega} \nonumber \\ \!\! &  & \!\!+\frac{1}{2} \left[ (\partial_\tau \vec{A}_1 + \nabla U_1)^2 - (\partial_\tau \vec{A}_2 + \nabla U_2)^2 + (\nabla \times \vec{A}_1)^2 - (\nabla \times \vec{A}_2)^2 \right] \nonumber \\ \!\! &  & \!\!- ij \left[ (\partial_\tau \vec{A}_1 + \nabla U_1) \cdot (\partial_\tau \vec{A}_2 + \nabla U_2) + (\nabla \times \vec{A}_1) \cdot (\nabla \times \vec{A}_2) \right] \nonumber \\ \!\! &  & \!\! + \frac{m^2}{2}  \Omega \overline{\Omega} + \frac{\lambda}{4!} (\Omega \overline{\Omega})^2 \Biggr\},
\end{eqnarray}
where, from the gauge fields, we have identified the quantities $\mathcal{A}_0$ and $\mathcal{B}_0$ with the potentials $U_1$ and $U_2$, as well as the vector $\vec{A}_1$ and $\vec{A}_2$ quantities associated with the components $\mathcal{A}^i$ and $\mathcal{B}^i$, respectively. Considering the case where the electric potentials are zero and neglecting temporal variations in the fields $\vec{A}_1$ and $\vec{A}_2$, the functional reduces to: 
\begin{eqnarray}
    F[\Omega, \bar{\Omega}, \vec{A}_1, \vec{A}_1] \!\! & = & \!\! \int d^3 x \mathcal{F}_{GL} \nonumber \\ \!\! & =& \!\! \int d^3 x \Biggl\{ [\nabla - ie(\vec{A}_1 - ij \vec{A}_2)] \Omega \cdot [\nabla + ie(\vec{A}_1 - ij \vec{A}_2)] \overline{\Omega} \nonumber \\ \!\! & & \!\! + \frac{1}{2} [\nabla \times( \vec{A}_1-ij\vec{A}_2) ]^2 + \frac{m^2}{2}  \Omega \overline{\Omega} + \frac{\lambda}{4!} (\Omega \overline{\Omega})^2 \Biggr\}, \label{Functional1}
\end{eqnarray}
 and by construction, the free energy density given by the functional $\mathcal{F}_{GL}$ is a Hermitian quantity and can be expressed in terms of two real functionals of the form:
 \begin{equation}
\mathcal{F}_{GL}=\mathcal{F}_{R}+ij\mathcal{F}_{H}.
\end{equation}
In order to dertermine explicitly the form of these functionals, we expand the first and fourth terms in the expression \eqref{Functional1}, and taking into account the explicit form of the hypercomplex field $\Omega$ and its norm of \eqref{norm1} we have that:

\begin{eqnarray}
    [\nabla - ie\vec{C}_H] \Omega \cdot [\nabla + ie\vec{C}_H] \overline{\Omega} \!\! & = & \!\! \Biggl\{ (\gamma^2-1)(\nabla \phi_2^2 + \nabla \phi_1^2) + (\gamma^2-1)e^2 (\vec{A}_1^2 - \vec{A}_2^2)(\phi_1^2 + \phi_2^2)\nonumber \\ \!\! & & \!\! + 4\gamma e^2 \vec{A}_1 \cdot \vec{A}_2 (\phi_2^2 - \phi_1^2)  + 2e(\gamma^2+1) \vec{A}_1 \cdot (\phi_2 \nabla \phi_1 - \phi_1 \nabla \phi_2) \Biggr\} \nonumber \\ \!\! & & \!\! + ij \Biggl\{ 2\gamma (\nabla \phi_2^2 - \nabla \phi_1^2)  + 2\gamma e^2 (\vec{A}_1^{2} - \vec{A}_2^2) \nonumber \\ \!\! & & \!\! - 2e(\gamma^2+1) \vec{A}_2 \cdot (\phi_2 \nabla \phi_1 - \phi_1 \nabla \phi_2) \Biggr\},
\label{term1}
\end{eqnarray}
where $\vec{C}_H = \vec{A}_1-ij\vec{A}_2$;
while the terms of the potential are expressed as follows:
\begin{eqnarray}
    V(\Omega, \overline{\Omega}) \!\! & = & \!\! \frac{m_R^2}{2} (\gamma^2 - 1)(\phi_1^2 + \phi_2^2) + \frac{\lambda_R}{4!} (\gamma^2 - 1)^2 (\phi_1^2 + \phi_2^2)^2 - \frac{\gamma^2 \lambda_R}{6} (\phi_2^2 - \phi_1^2)^2 \nonumber \\ \!\! & & \!\! - m_H^2 \gamma (\phi_2^2 - \phi_1^2) - \frac{\lambda_H}{6} \gamma (\gamma^2- 1) (\phi_2^4 - \phi_1^4) + ij \Biggl\{m_R^2 \gamma (\phi_2^2 - \phi_1^2) +\nonumber \\ \!\! & & \!\! \frac{\lambda_R}{6} \gamma (\gamma^2 - 1) (\phi_2^4 - \phi_1^4)  + \frac{ m_H^2}{2} (\gamma^2 - 1) (\phi_1^2 + \phi_2^2)\nonumber \\ \!\! & & \!\! + \frac{\lambda_H}{4!} (\gamma^2 - 1)^2 (\phi_1^2 + \phi_2^2)^2 - \frac{\lambda_H \gamma^2}{6} (\phi_2^2 - \phi_1^2)^2 \Biggr\}.
\label{term2}
\end{eqnarray}
Note that it has the form $V=V_R+ijV_R$ in which $V_R$ and $V_H$ are associated with the subsystem and reservoir respectively. 
From \eqref{term1} and \eqref{term2}, we can then define the functional
\begin{eqnarray}
  \mathcal{F}_{R} \!\! & = & \!\! (\gamma^2-1)(\nabla \phi_2^2 + \nabla \phi_1^2) + (\gamma^2-1)e^2 (\vec{A}_1^2 - \vec{A}_2^2)(\phi_1^2 + \phi_2^2)
     + 4\gamma e^2 \vec{A}_1 \cdot \vec{A}_2 (\phi_2^2 - \phi_1^2) \nonumber \\ \!\! & & \!\! + 2e(\gamma^2+1) \vec{A}_1 \cdot (\phi_2 \nabla \phi_1 - \phi_1 \nabla \phi_2)  + \frac{1}{2} [(\nabla \times \vec{A}_1)^2 - (\nabla \times \vec{A}_2)^2 ]+a_{1}(\phi_1^2+\phi_2^2) \nonumber \\ \!\! & & \!\! +a_{2}(\phi_2^2-\phi_1^2) +b_{1}(\phi_1^2+\phi_2^2)^2+b_{2}(\phi_2^2-\phi_1^2)^2+b_{3}(\phi_2^4-\phi_1^4),
     \label{fr}
\end{eqnarray}
corresponding to the subsystem, and
\begin{eqnarray}
    \mathcal{F}_{H} \!\! & = & \!\! 2\gamma (\nabla \phi_2^2 - \nabla \phi_1^2) + 2\gamma e^2 (\vec{A}_1^2 - \vec{A}_2^2) - 2e(\gamma^2+1) \vec{A}_2 \cdot (\phi_2 \nabla \phi_1 - \phi_1 \nabla \phi_2) \nonumber \\ \!\! & & \!\! -(\nabla \times \vec{A}_1) \cdot (\nabla \times \vec{A}_2)+c_{1}(\phi_1^2+\phi_2^2)+c_{2}(\phi_2^2-\phi_1^2)+d_{1}(\phi_1^2+\phi_2^2)^2 \nonumber \\ \!\! & & \!\! +d_{2}(\phi_1^2-\phi_2^2)^2+d_{3}(\phi_2^4-\phi_1^4),
\label{fh} 
\end{eqnarray}
for the reservoir. Where the coefficient are defined as:
\begin{eqnarray}
     a_{1} \!\! & = & \!\! m_{R}^{2}\frac{(\gamma^2-1)}{2}, \quad a_{2}=-\gamma  m_{H}^{2}, \quad b_{1} =\frac{\lambda_{R}}{4!}(\gamma^2-1)^2, \nonumber \\ b_{2}\!\! & =& \!\! \frac{-\lambda_{R}}{6}\gamma^2, \quad b_{3} =\frac{-\lambda_{H}}{6}\gamma(\gamma^2-1).
           \label{coefR}
\end{eqnarray}
\begin{equation}
   \begin{split}
       c_{1} &= \frac{m_{H}^{2}}{2}(\gamma^2-1), \quad c_{2}=m_{R}^{2}\gamma, \quad d_{1}=\frac{\lambda_{H}}{4!}(\gamma^2-1)^2, \\ d_{2} & =\frac{-\lambda_{H}}{6}\gamma^2, \quad d_{3}= \lambda_{R}\frac{\gamma(\gamma^2-1)}{6}. \\ 
       \label{coefH}
   \end{split}
\end{equation}
\subsection{The Ginzburg-Landau equations}
\label{G-L}
To determine the stable configurations of the system, it is essential to establish the macroscopic equilibrium conditions of the model. Analytically, this is achieved by requiring that the free energy functional takes on an extreme value, which ensures that the steady state corresponds to a configuration of minimum energy. Consequently, the formal procedure requires applying the variational principle to the Ginzburg–Landau functional. First, we calculate the total variation of expression \eqref{Functional1}, obtaining:
\begin{eqnarray}
    \delta F[\Omega, \overline{\Omega}, \vec{C}_H] \!\! & = &\!\! \delta \int d^3x [\nabla - ie\vec{C}_H]\Omega \cdot [\nabla + ie\vec{C}_H]\overline{\Omega} \nonumber \\ \!\! & & \!\! + \delta \int \frac{1}{2}[\nabla \times \vec{C}_H]^2 d^3x + \delta \int d^3x \left[ \frac{m^2}{2} \Omega \overline{\Omega} + \frac{\lambda}{4!}(\Omega \overline{\Omega})^2 \right] \nonumber \\ \!\! &= & \!\! \int d^3x \Biggl\{ \left[ \frac{m^2}{2} \Omega + \frac{\lambda}{12}(\Omega \overline{\Omega})\Omega+(-i\nabla -e\vec{C}_H)^2 \Omega \right] \delta \overline{\Omega} + c.c \nonumber \\ \!\! & & \!\! + \left[ \nabla \times (\nabla \times \vec{C}_H) + ie [ \bar{\Omega}\nabla\Omega-\Omega\nabla \overline{\Omega}]+2e^2\Omega\overline{\Omega}\vec{C}_H \right] \cdot \delta \vec{C}_H \Biggr\} \nonumber \\ \!\! & & \!\! + \oint \Biggl\{ [(\nabla - ie\vec{C}_H)\Omega \cdot \hat{n}] \delta \overline{\Omega} + [(\nabla \times \vec{C}_H) \times \hat{n}] \cdot \delta \vec{C}_H \Biggr\} dS.
\label{TotalV}
\end{eqnarray}
From this result, we deduce the two equilibrium conditions by simultaneously requiring that the free energy attains a minimum with respect to the conjugate order parameter $\frac{\delta F}{\delta \overline{\Omega}} = 0$, and with respect to the generalized potential vector, $\frac{\delta F}{\delta \vec{C}_H} = 0$:
\begin{equation}
    (-i\nabla -e\vec{C}_H)^2 \Omega+\frac{m^2}{2} \Omega + \frac{\lambda}{12}(\Omega \bar{\Omega})\Omega=0,
\label{GL1}
\end{equation}
\begin{equation}
    \nabla \times (\nabla \times \vec{C}_H) =-ie [ \bar{\Omega}\nabla\Omega-\Omega\nabla \bar{\Omega}]-2e^2\Omega\bar{\Omega}\vec{C}_H.
\label{GL2}
\end{equation}
If we expand equations \eqref{GL1} in terms of the components $\phi_1$, $\phi_2$, $\vec{A}_1$ and $\vec{A}_2$, using the real and hyperbolic parts, we obtain the following pair of equations:
\begin{eqnarray}
    \!\! & & \!\! -\nabla^2\phi_2 + e^2(\vec{A}_1^2 - \vec{A}_2^2)\phi_2 - \left(\frac{\gamma^2-1}{\gamma^2+1}\right)e(\nabla\cdot\vec{A}_1 + 2\vec{A}_1\cdot\nabla)\phi_1 + \left(\frac{2\gamma}{\gamma^2+1}\right)e(\nabla\cdot\vec{A}_2 + 2\vec{A}_2\cdot\nabla)\phi_1 \nonumber \\ \!\! & & \!\! + \frac{m_R^2}{2}\phi_2  + \frac{\lambda_R}{12}( \gamma^2-1)(\phi_2^2+\phi_1^2)\phi_2+\frac{\gamma\lambda_H}{6}(\phi_1^2-\phi_2^2)\phi_2 = 0,
\label{GL3}
\end{eqnarray}
\begin{equation}
    \begin{split}
        &-(\gamma^2-1)\nabla^2\phi_1 + e^2(\gamma^2-1)(\vec{A}_1^2 - \vec{A}_2^2)\phi_1 - 4\gamma e^2(\vec{A}_1\cdot\vec{A}_2)\phi_1 + (\gamma^2+1)e(\nabla\cdot\vec{A}_1 + 2\vec{A}_1\cdot\nabla)\phi_2 \\ & + \frac{1}{2} \left[ m_R^2(\gamma^2-1) + 2\gamma m_H^2 \right] \phi_1 +  \left[ \frac{\lambda_R}{12} (\gamma^2-1)^2(\phi_2^2+\phi_1^2) + \frac{\lambda_R}{6}\gamma^2 (\phi_2^2-\phi_1^2) + \frac{\lambda_H}{3} \gamma   (\gamma^2-1) \phi_1 \right] \phi_1 = 0;
    \end{split}
\label{GL4}
\end{equation}
and similarly for \eqref{GL2}:
\begin{eqnarray}
\nabla \times (\nabla \times \vec{A}_1) =2e(\gamma^2+1)[ \phi_2\nabla\phi_1-\phi_1\nabla \phi_2] - 2e^2 [(\gamma^2-1)(\phi_2^2+\phi_1^2)\vec{A}_1 +\gamma (\phi_2^2-\phi_1^2)\vec{A}_2],  
\label{GL5}
\end{eqnarray}
\begin{equation}
    \nabla \times (\nabla \times \vec{A}_2)=2e^2[2\gamma(\phi_2^2-\phi_1^2)\vec{A}_1-(\gamma^2-1)(\phi_2^2+\phi_1^2)\vec{A}_2].
    \label{GL6}
\end{equation}
The expressions \eqref{GL3} to \eqref{GL6} constitute the generalized Ginzburg–Landau equations for hypercomplex scalar electrodynamics. These equations govern the dynamic of the matter fields, the penetration profile of the interactions, and the distribution of the currents associated with thermodynamic equilibrium for a dissipative system constituted by a sub-system and reservoir.
\subsection{Free energy functional in the vacuum state}
\label{vacuum}
Given the underlying symmetry, the construction of a phenomenological theory begins with an evaluation of the allowed invariants under $U(1)\times SO(1,1)$. First, we assume a vacuum scenario in which there are no magnetic fields and a homogeneous condition for the fields; the energy densities for the real sector and the dual sector of expressions \eqref{fr} and \eqref{fh} are reduced to the following forms:
\begin{equation}
\begin{split}
\mathcal{F}_{R} &=
 a_{1}\mathfrak{I}_1 +a_{2} \mathfrak{I}_2 
 + b_{1}\mathfrak{I}_1^2 + b_{2}\mathfrak{I}_2^2 + b_{3}\mathfrak{I}_1\mathfrak{I}_2,
 \label{FRvac}
\end{split}
\end{equation}
\begin{equation}
        \begin{split}
\mathcal{F}_{H} &= c_{1}\mathfrak{I}_1 + c_{2}\mathfrak{I}_2 + d_{1}\mathfrak{I}_1^2 + d_{2}\mathfrak{I}_2^2 + d_{3}\mathfrak{I}_1\mathfrak{I}_2,
\label{FHvac}
\end{split}
\end{equation}
where we have defined the circular and hyperbolic invariants: $\mathfrak{I}_1=(\phi_1^2+\phi_2^2)$, $\mathfrak{I}_2 =(\phi_2^2-\phi_1^2)$, and the associated coefficients are given by expressions \eqref{coefR} and \eqref{coefH}.
The coefficients for the quadratic terms govern the curvature of the potential, while the quartic terms $b_i$ and $d_i$ determine the fundamental rules of self-interaction; thus, the evolution of the functionals is determined by the parameter $\gamma$. Previous works have shown that the introduction of a hyperbolic gauge symmetry induces profound and non-trivial effects on the topology of the vacuum \cite{ssbh,Hyper4}. Furthermore, under this same formalism, the symmetry-breaking scenarios studied suggest that $\gamma$ is not merely a constant that influences the geometry, but can also be regarded as a fundamental phenomenological parameter that governs the dissipative and directional behavior of the coupled system. On the other hand, in the Ginzburg-Landau theory of phase transitions in the mean-field regime, the symmetry breaking is governed by the sign change of the isotropic mass term, which  exhibits linear thermal dependence. In our model, will show that it is the hyperbolic contribution that dictates the critical behavior and the reconfiguration of the vacuum.  
Therefore, following Landau’s ideas \cite{Landau1958,Landau1936,goldenfeld}, we propose that the coefficients associated with the hyperbolic norm invariants $a_2$ and $c_2$ can be expressed as a series expansion of the temperature near a critical point of the form:
\begin{eqnarray}
    a_2\!\! & = & \!\!\alpha_0+\alpha(T-T_c)+O((T-T_c)^2), \nonumber \\ c_2\!\! & = &\beta_0+\beta(T-T_c)+O((T-T_c)^2);
\end{eqnarray}
considering a first-order approximation and that the coefficient of the hyperbolic invariant is determined directly by comparing with the expression \eqref{coefR}, where $a_{2}=-\gamma  m_{H}^{2}$, then this constraint leads us to the form for $\gamma$:
\begin{equation}
\gamma(T)=1-\frac{T}{T_c}.
\label{gammaT}
\end{equation}
 When the thermal dependence of $\gamma(T)$ is incorporated into the functionals, the coefficients of the quadratic terms in the real sector take the form:
\begin{equation}
    a_1(T)=\frac{m_R^2}{2}\left[\left(1- \frac{T}{T_c} \right)^2-1  \right], \quad \quad \quad a_2(T)=-m_H^2 \left( 1-\frac{T}{T_c}\right),
    \label{ai}
\end{equation}
and similarly for the hybrid sector:
\begin{equation}
    c_1(T)=\frac{m_H^2}{2}\left[\left(1- \frac{T}{T_c} \right)^2-1  \right], \quad \quad \quad c_2(T)=m_R^2 \left( 1-\frac{T}{T_c}\right).
\end{equation}
We can find the energy minima or the most probable configuration of the system using expression \eqref{GL1}, which in the vacuum state reduces to:
\begin{equation}
    \frac{m^2}{2} \Omega + \frac{\lambda}{12}(\Omega \bar{\Omega})\Omega=0,
\end{equation}
this leads to stable minima for the pair:
\begin{eqnarray}
    \left\langle \phi_{1} \right\rangle_0^2 \!\! & = & \!\! \frac{c_1(2b_1+b_3)-a_1(2d_1+d_3)}{4(b_1d_3-b_3d_1)}, \nonumber \\ 
    \left\langle \phi_{2} \right\rangle_0^2 \!\! & = & \!\! \frac{a_1(2d_1-d_3)-c_1(2b_1-b_3)}{4(b_1d_3-b_3d_1)}.
    \label{min1}
\end{eqnarray}
A direct analysis of these analytical expressions demonstrates a singular behavior induced by the interplay between the quadratic coefficients and the quartic couplings near the critical point $T = T_c$ ($\gamma = 0$), precisely where the invariant norm of the hypercomplex field in \eqref{norm1} becomes purely real and restricts the system to exclusively circular rotations. Conversely, at the limits of maximal coupling ($T \to 0$ or $T = 2T_c$, where $\gamma^2 = 1$), the norm reduces entirely to its hyperbolic sector, which completely suppresses circular rotations and renders the vacuum expectation values in \eqref{min1} indeterminate. This suggests that the system undergoes a spontaneous directional symmetry breaking and is forced to abandon any mixed-state configuration, so the uncertainty must be resolved by strictly choosing a vacuum with preferential direction; as a result, the dynamics collapse into a single accessible mode, and by adopting the configuration where $\langle \phi_{2} \rangle_0^2 = 0$, one has that:
\begin{equation}
    \frac{c_1}{a_1}=\frac{2d_1-d_3}{2b_1-b_3}=\frac{m_H^2}{m_R^2}, \qquad \qquad \left\langle \phi_{1} \right\rangle_0^2=\frac{a_1}{(b_3-2b_1)}=\frac{6m_R^2}{[1-(1-\frac{T}{T_c})^2]\lambda_R-2\lambda_H(1-\frac{T}{T_c})};
    \label{conf1}
\end{equation}
or alternatively $ \left\langle \phi_{1} \right\rangle_0^2=0$:
\begin{equation}
    \frac{c_1}{a_1}=\frac{2d_1+d_3}{2b_1+b_3}=\frac{m_H^2}{m_R^2}, \qquad \qquad \left\langle \phi_{2} \right\rangle_0^2=\frac{a_1}{(2b_1+b_3)}=\frac{6m_R^2}{[1-(1-\frac{T}{T_c})^2]\lambda_R+2\lambda_H(1-\frac{T}{T_c})};
    \label{conf2}
\end{equation}
as it has been well illustrated in \cite{ssbh}.

\subsubsection{Changes in the geometric structures of the vacuum manifold induced by the instability of self-interaction constants}
\label{qedvacuum2} 
\begin{figure}[htbp]
    \centering
    \begin{subfigure}[b]{0.32\textwidth}
        \centering
        \includegraphics[width=\textwidth]{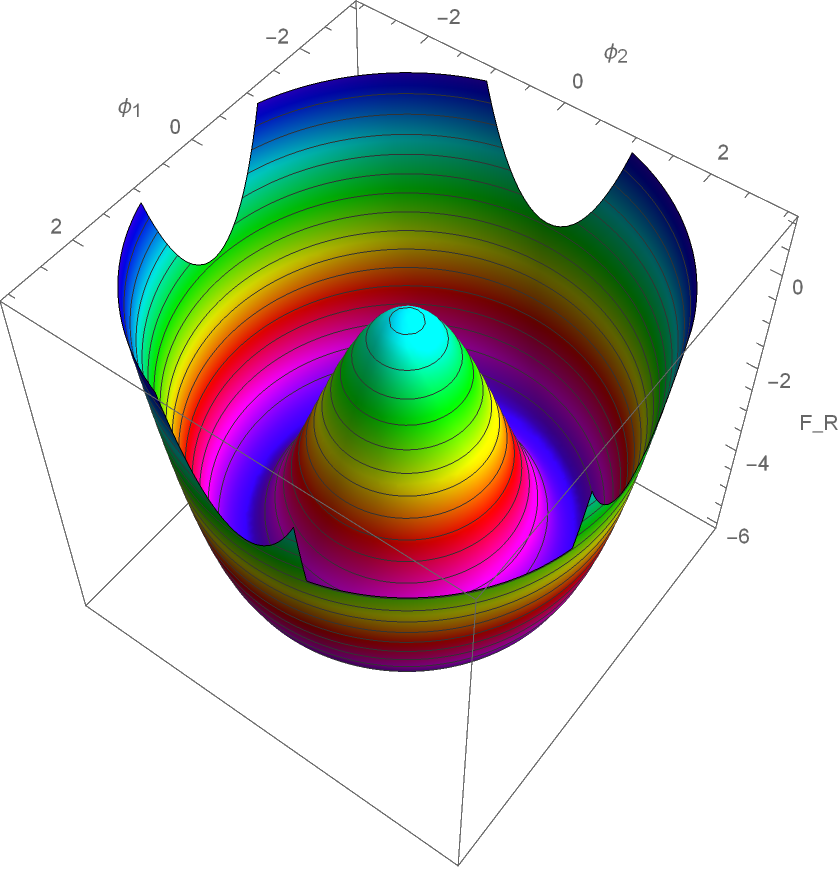}
        \caption{}
        \label{fig:1a}
    \end{subfigure}
    \hfill
    \begin{subfigure}[b]{0.32\textwidth}
        \centering
        \includegraphics[width=\textwidth]{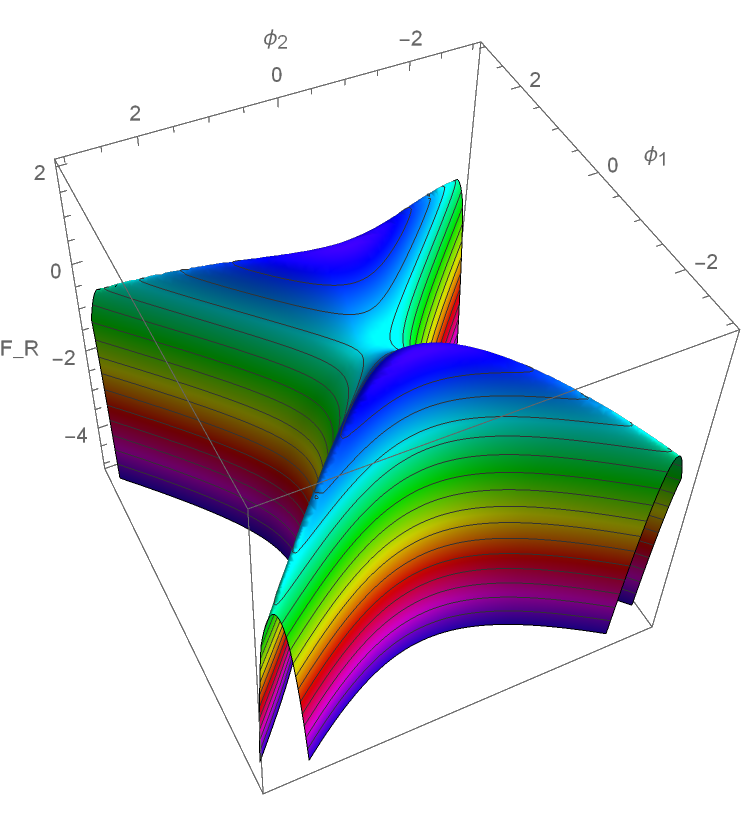}
        \caption{}
        \label{fig:1b}
    \end{subfigure}
    \hfill
    \begin{subfigure}[b]{0.32\textwidth}
        \centering
        \includegraphics[width=\textwidth]{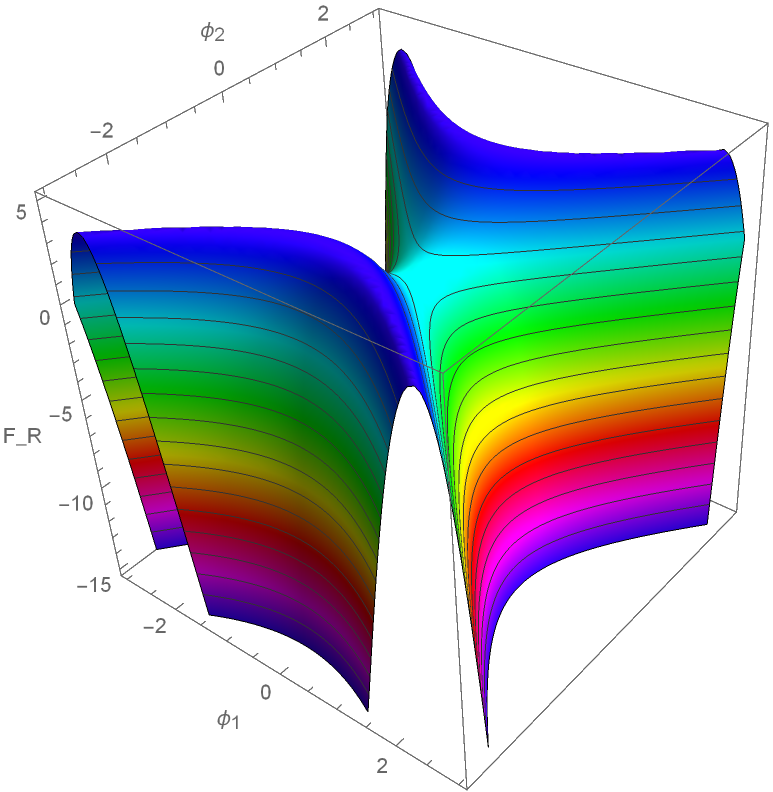}
        \caption{}
        \label{fig:1c}
    \end{subfigure}
    \caption{Free-energy density landscape for the illustrative positive coupling scenario $\lambda_R, \lambda_H > 0$: a) the symmetric Mexican-hat potential restored at $T=T_c$ ($\gamma=0$) for $\mathcal{F}_R$ and $\mathcal{F}_R$, the directional destabilization induced by hyperbolic invariants as $T \to 0$ ($\gamma\to 1$) for the functionals b) $\mathcal{F}_R$ and c) $\mathcal{F}_H$.}
    \label{Fig1}
\end{figure}

\begin{figure}[ht]
    \centering
    \begin{subfigure}[b]{0.30\textwidth}
        \centering
        \includegraphics[width=\textwidth, height=4cm]{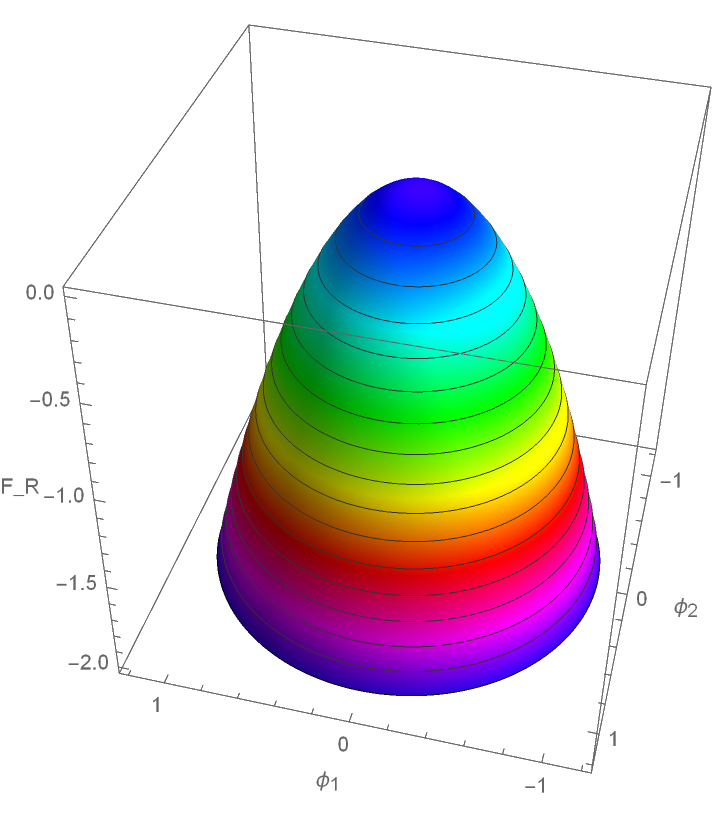}
        \caption{}
        \label{fig:2a}
    \end{subfigure}
    \hspace{2pt}
    \begin{subfigure}[b]{0.37\textwidth}
        \centering
        \includegraphics[width=\textwidth, height=4cm]{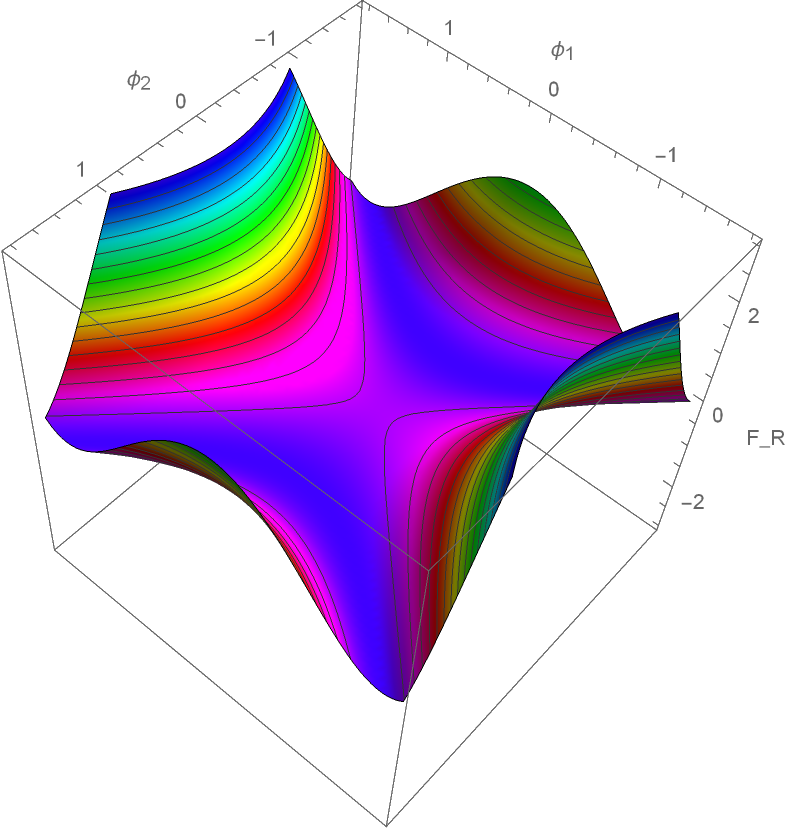}
        \caption{}
        \label{fig:2b}
    \end{subfigure}
    
    \vspace{1ex} 
    
    \begin{subfigure}[b]{0.32\textwidth}
        \centering
        \includegraphics[width=\textwidth, height=4cm]{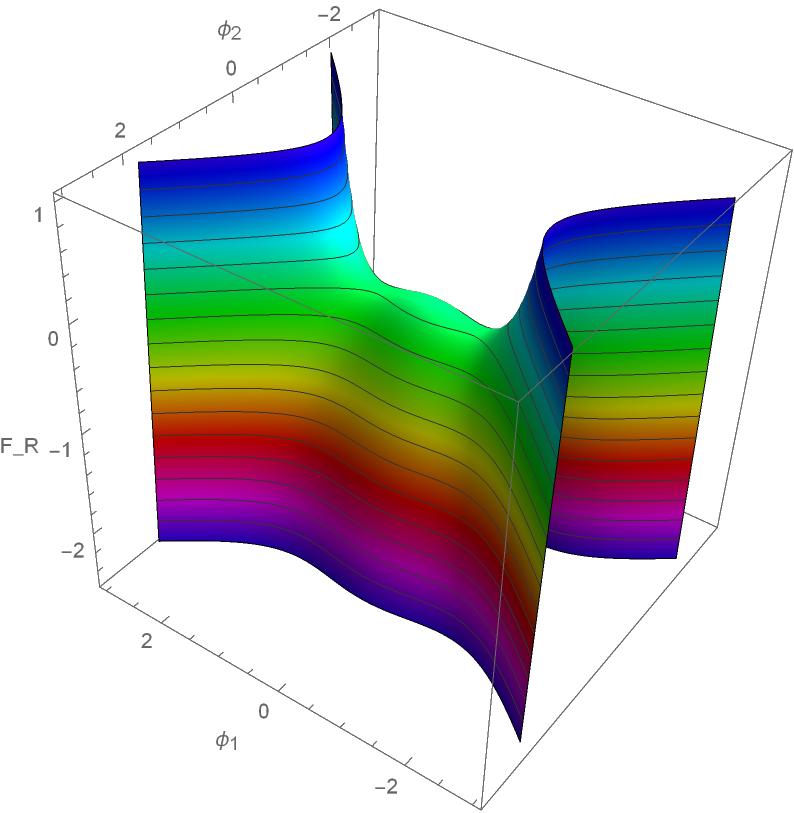}
        \caption{}
        \label{fig:2c}
    \end{subfigure}
    \hspace{2pt}
    \begin{subfigure}[b]{0.34\textwidth}
        \centering
        \includegraphics[width=\textwidth, height=4cm]{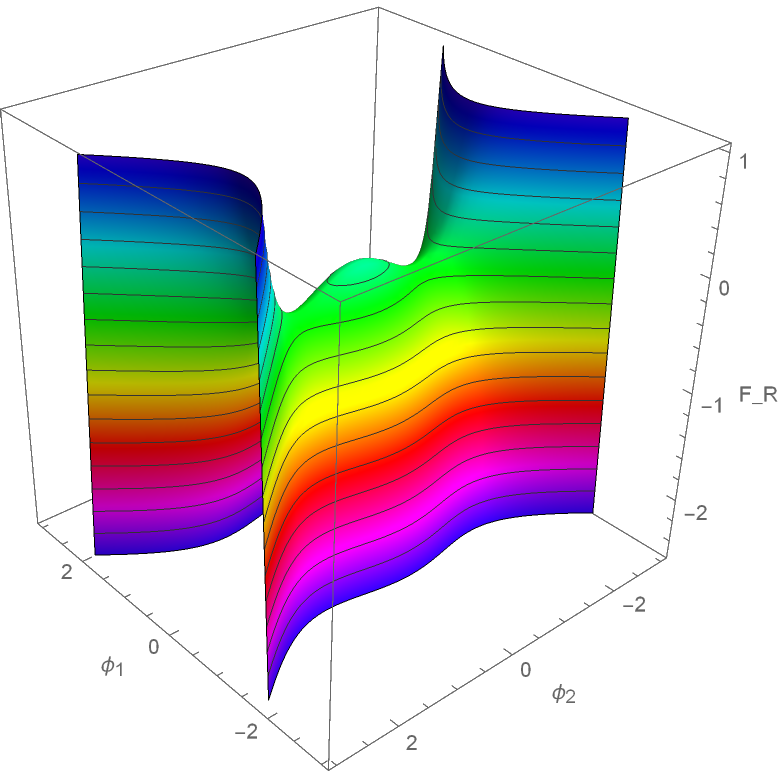}
        \caption{}
        \label{fig:2d}
    \end{subfigure}
     \caption{Free energy density $\mathcal{F}_R$ for the case of negative coupling ($\lambda_R, \lambda_H < 0$) showing a) the inverted surface at $T=T_c$, b) the non-compact hyperbolic valleys as $T \to 0$ unbounded from below in the intermediate regimes c) $0<T<T_c$ and d) $T>T_c$).}
    \label{fig2}
\end{figure}

\begin{figure}[H]
    \centering
    \begin{subfigure}[b]{0.30\textwidth}
        \centering
        \includegraphics[width=\textwidth, height=4cm]{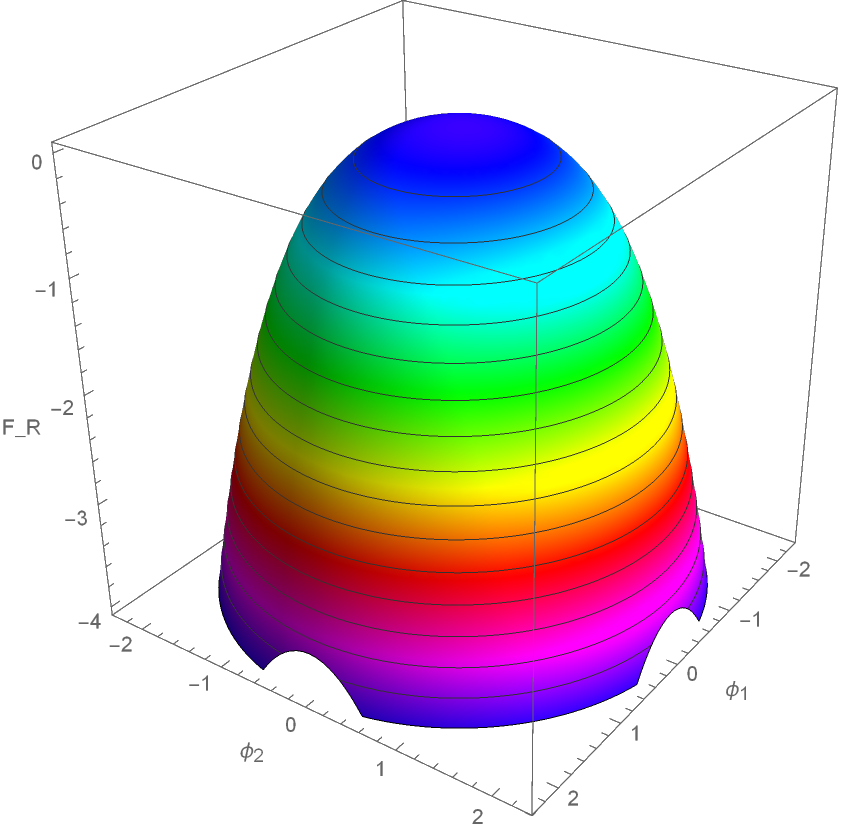}
        \caption{}
        \label{fig:3a}
    \end{subfigure}
    \hspace{2pt}
    \begin{subfigure}[b]{0.37\textwidth}
        \centering
        \includegraphics[width=\textwidth, height=4cm]{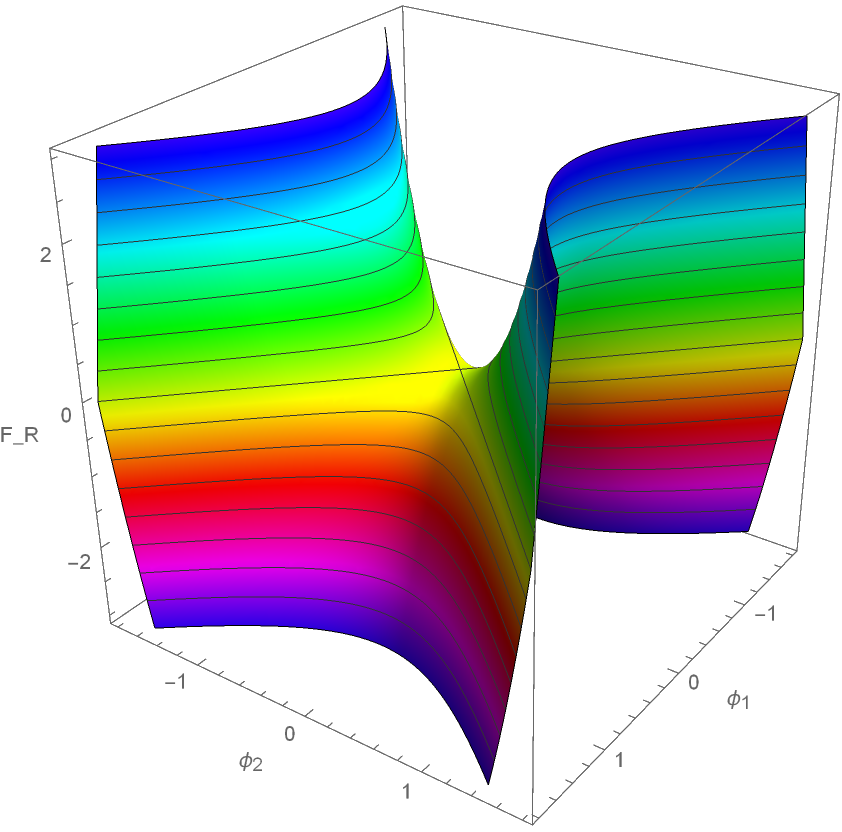}
        \caption{}
        \label{fig:3b}
    \end{subfigure}
    
    \vspace{1ex} 
    
    \begin{subfigure}[b]{0.32\textwidth}
        \centering
        \includegraphics[width=\textwidth, height=4cm]{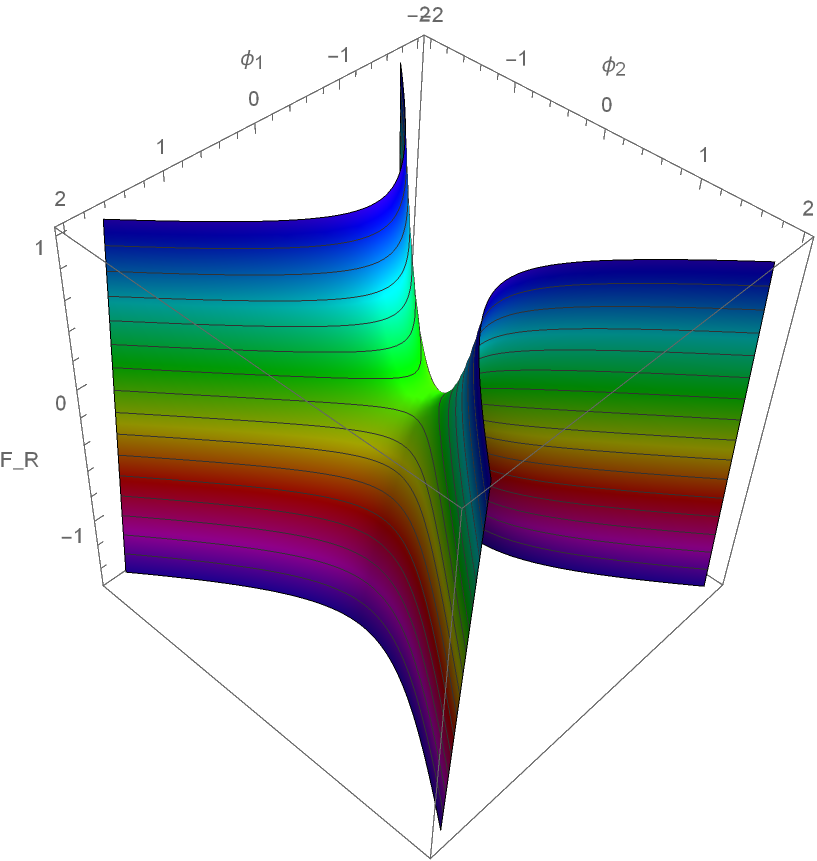}
        \caption{}
        \label{fig:3c}
    \end{subfigure}
    \hspace{2pt}
    \begin{subfigure}[b]{0.34\textwidth}
        \centering
        \includegraphics[width=\textwidth, height=4cm]{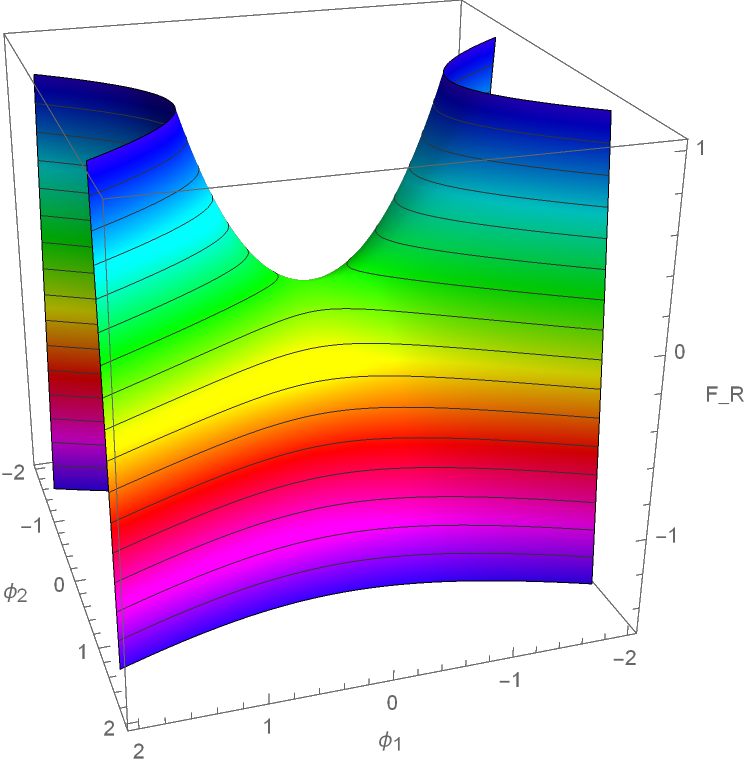}
        \caption{}
        \label{fig:3d}
    \end{subfigure}
     \caption{Free energy density $\mathcal{F}_H$ for the case of negative coupling ($\lambda_R, \lambda_H < 0$) at a) $T=T_c$, b)$T \to 0$, c) $0<T<T_c$ and d) $T>T_c$).}
    \label{fig3}
\end{figure}

The variation in vacuum geometric is intrinsically determined by the nature of the fundamental self-interaction constants, $\lambda_R$ and $\lambda_H$. The free-energy landscape at the critical point $T=T_c$ reveals a dichotomy dictated by the sign of these parameters,
if we impose positive constants ($\lambda_R, \lambda_H > 0$), the quartic coefficients will guarantee a lower bound for the free energy densities in the functionals $\mathcal{F}_R$ and $\mathcal{F}_H$ near the critical point. In this illustrative scenario, when the temperature crosses exactly at $T = T_c$, the functionals momentarily recovers the $U(1)$ symmetry, and its geometric shape will be the usual Mexican hat, as shown in Fig.1(\subref{fig:1a}). On the other hand, in a temperature regime close to absolute zero, the system is governed by hyperbolic norm invariants. In this case the accompanying quartic coefficients undergo a sign change that makes the free energy density of both $\mathcal{F}_R$ and $\mathcal{F}_H$ are completely unbounded and unstable along specific directions, as illustrated in Fig.1(\subref{fig:1b}) and (\subref{fig:1c}). 
On the contrary, when $\lambda_R, \lambda_H < 0$, the vacuum configuration undergoes profound structural collapses. At the critical temperature $T=T_c$, the negativity of the effective quartic coefficients completely eliminates the confining barriers, the Mexican hat thus transforms into an inverted and unstable surface; within this configuration the fields do not encounter restoring forces, and the solutions diverge asymptotically toward $-\infty$ along the axes of the field components as illustrated in the Fig.2(\subref{fig:2a}) and Fig.3(\subref{fig:3a}). As the system cools to temperatures very close to absolute zero ($T \to 0$), the geometry transits to a hyperbolic profile configuration. Here, in the case of the real sector $\mathcal{F}_R$ the effective quadratic and quartic coefficients take on positive values along a specific direction, creating a saddle-like geometry characterized by two distinct non-compact vacuum state valleys that extend along the preferred axis $\phi_2$, so in this regimen is momentarily governed by the symmetry $SO(1,1)$. A similar behavior manifests within the hybrid sector $\mathcal{F}_H$ (see eq.\eqref{FHvac}), with the crucial distinction that the hyperbolic symmetry $SO(1,1)$ is not recovered in the low-temperature limit since the quadratic coupling in this sector remains positive. 
Fig.2(\subref{fig:2b}) and Fig.3 (\subref{fig:3b}) shows the free energy densities for $\mathcal{F}_R$ and $\mathcal{F}_H$  for the case of negative coupling.

\noindent Furthermore, the Fig.2(\subref{fig:2c}),(\subref{fig:2d}) and the Fig.3(\subref{fig:3c}),(\subref{fig:3d}) show intermediate thermal regimes, i.e., for $0 < T < T_c$ $(0<\gamma<0)$ and $T > T_c$ $(-1<\gamma<0)$, where there is no pure circular and hyperbolic symmetry, instead there is a non-trivial combination of the $U(1) \times SO(1,1)$ symmetry. Within these regions the non-zero thermal parameter $\gamma$ activates the directional mass and coupling terms in all sectors; however, due to the persistent negativity of the coefficients accompanying the fourth-order invariants $\mathfrak{I}_2$, this symmetry breaking does not stabilize the vacuum; instead, the landscape develops unbounded and highly asymmetric regions that slope abruptly towards negative infinity, highlighting a generalized thermodynamic instability at all intermediate temperatures.
\subsubsection{Directional competence and ordering of the fields}
\begin{figure}[H]
    \centering
    \begin{subfigure}[b]{0.30\textwidth}
        \centering
        \includegraphics[width=\textwidth, height=5cm]{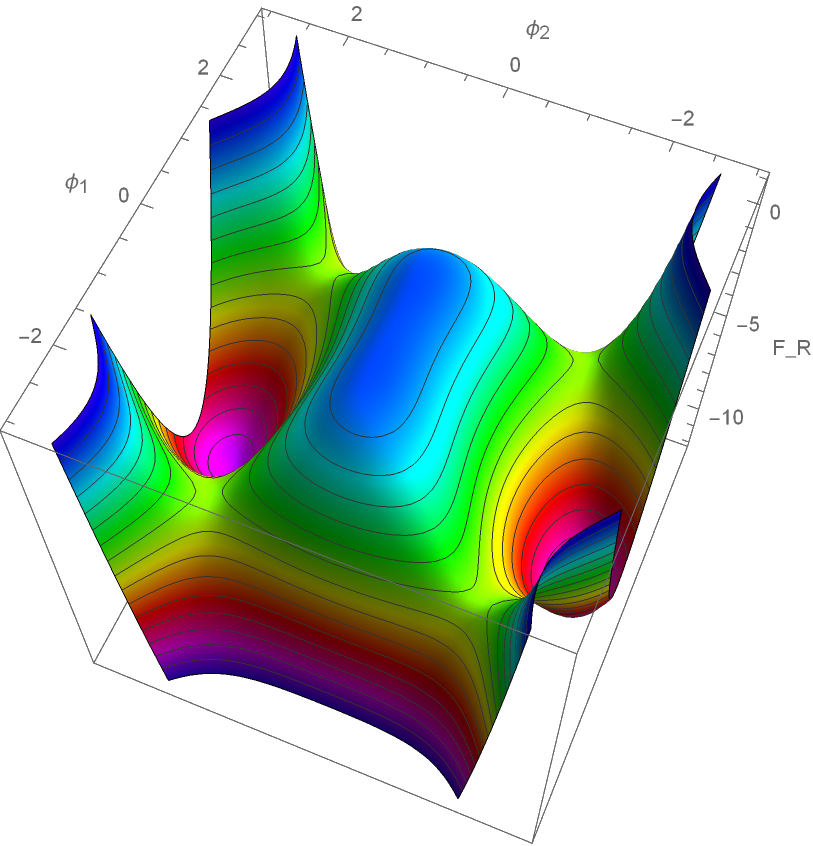}
        \caption{}
        \label{fig:4a}
    \end{subfigure}
    \hspace{2pt}
    \begin{subfigure}[b]{0.37\textwidth}
        \centering
        \includegraphics[width=\textwidth, height=5cm]{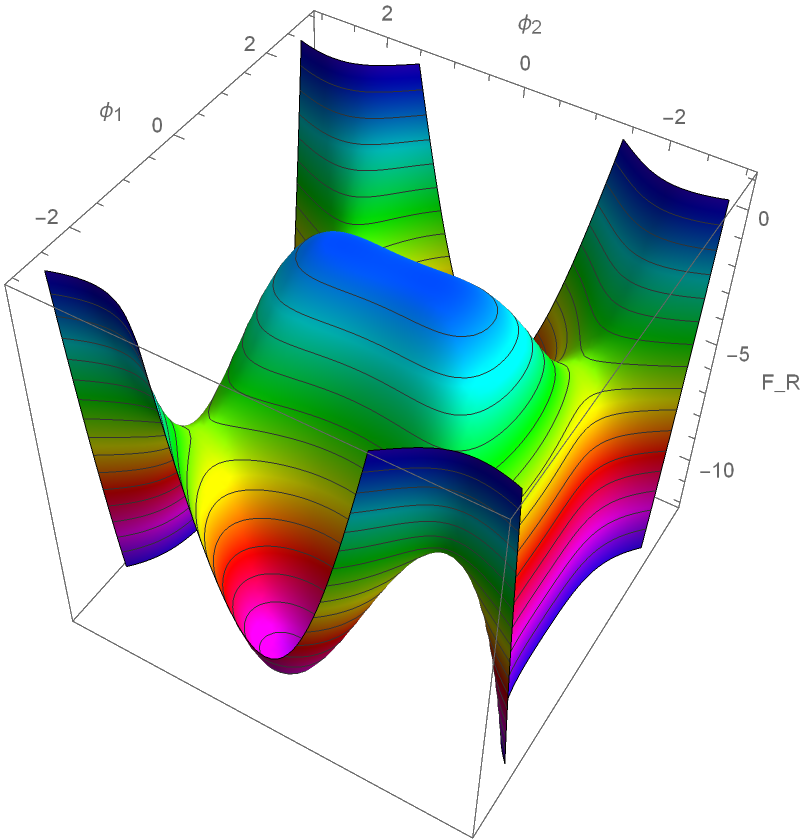}
        \caption{}
        \label{fig:4b}
    \end{subfigure}
    
    \vspace{1ex} 
    
    \begin{subfigure}[b]{0.32\textwidth}
        \centering
        \includegraphics[width=\textwidth, height=5cm]{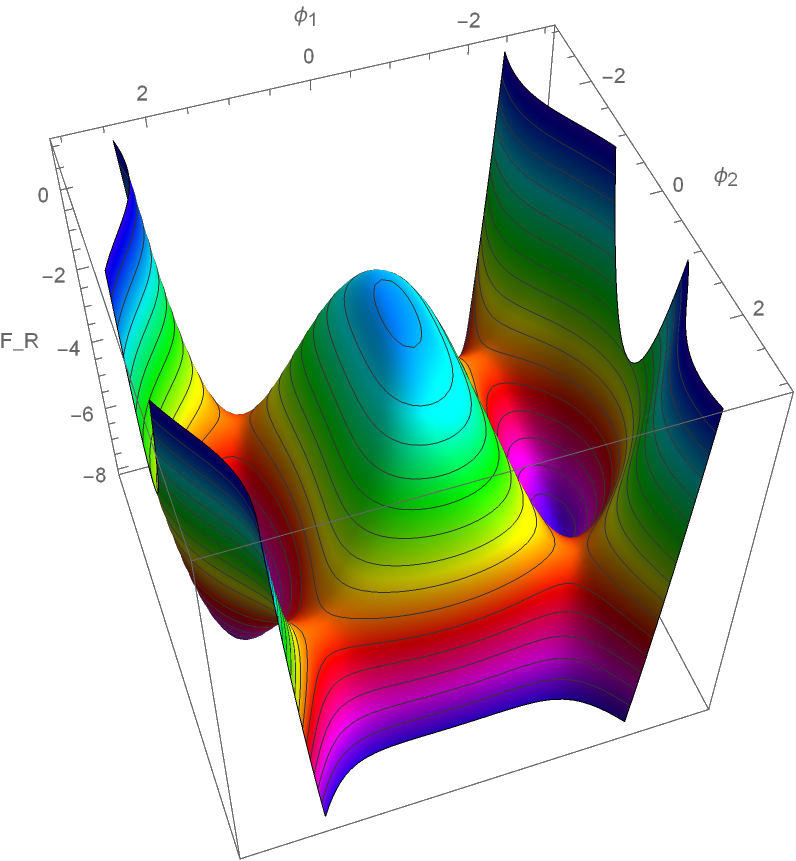}
        \caption{}
        \label{fig:4c}
    \end{subfigure}
    \hspace{2pt}
    \begin{subfigure}[b]{0.34\textwidth}
        \centering
        \includegraphics[width=\textwidth, height=5cm]{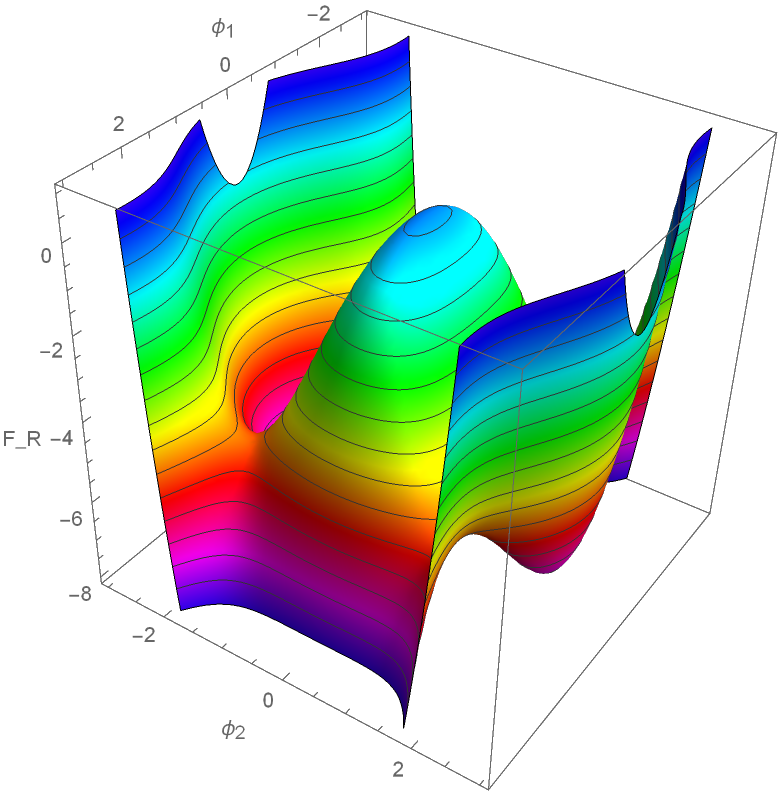}
        \caption{}
        \label{fig:4d}
    \end{subfigure}
     \caption{Free energy density $\mathcal{F}_R$ for a) $T=0.55T_c$ ($\gamma=0.45$) and b) $T=1.45T_c$ ($\gamma=-0.45$); free energy density $\mathcal{F}_H$ for c) $T=0.65T_c$ ($\gamma=0.35$) and d) $T=1.3T_c$ ($\gamma=-0.3$).}
    \label{fig4}
\end{figure}

As we have seen, the vacuum state can take on two configurations given by expressions \eqref{conf1} and \eqref{conf2}, which can be determined under certain temperature conditions. To do this, we define the effective masses associated with the field components $\phi_1$ and $\phi_2$, respectively. For the functional $\mathcal{F}_R$, these are:
\begin{equation}
    M_{R1}^2=a_1(T)-a_2(T); \quad  M_{R2}^2=a_1(T)+a_2(T). 
\end{equation}
\noindent Futhermore the competition between $a_{1}$ and $a_{2}$ determines which of these effective masses will have the greater influence, thereby dictating the system’s preferred direction of symmetry breaking, a similar hierarchy occurs with the hybrid part functional $\mathcal{F_H}$ via the effective masses $M_{H1}^2=c_1-c_2$ and  $M_{H2}^2=c_1+c_2$.
If we assume the stability condition where $\lambda_R=\lambda_H>0$, two main thermodynamic regimes are established:
\begin{enumerate}
    \item \textbf{Low-temperature regimen $0<T<T_c$ ($0<\gamma<1$):} In this region, the coefficients $a_1$ and $a_2$ are negative, and consequently $M_{R2}^2<M_{R1}^2$. Physically, we can say that the thermal bath forces the system to seek its minimum-energy state along the $\phi_2$ axis for the functional $\mathcal{F_R}$. This scenario is shown in Fig.4(\subref{fig:4a}), where the temperature reaches a value of $0.55T_c$ ($\gamma=0.45$). On the contrary, in the hybrid sector, it holds that $M_{H1}^2<M_{H2}^2$, and thus the minimum energy state of $\mathcal{F}_H$ corresponds to the axis of the $\phi_1$ field component, as shown in Fig.4(\subref{fig:4c}) with a value of $T=0.65T_c$ ($\gamma=0.35$).
    \item \textbf{High-temperature regimen $T>T_c$ ($\gamma <0$):} When the theoretical critical point $T_c$ is exceeded, $\gamma$ takes on negative values, and the coefficient of the invariant $\mathfrak{I_2}$ of $\mathcal{F}_R$ changes of sign ($a_2 > 0$), which alters the hierarchies, resulting in $M_{R1}^2 < M_{R2}^2$. This reversal causes the system to shift its direction from the stable minima toward the $\phi_1$ axis. The Fig.4(\subref{fig:4b}) illustrates this case for a temperature of $1.45T_c$ which correspond to $\gamma=-0.45$. For the functional $\mathcal{F}_H$, it now occurs that $M_{H2}^2<M_{H2}^2$, and the preferential axis for the stable state is now located at $\phi_2$. This situation is represented by Fig.4(\subref{fig:4d}) for a value of $T=1.3T_c$ ($\gamma=-0.3$).
\end{enumerate}
The Landau phenomenology for two-parameter coupled systems—the $O(N)$ broken symmetry is observed in anisotropic Heisenberg models applied to metamagnets \cite{birgeneau, rossini}; for this system the nature of the critical point is governed by the discriminant $\Delta = u_1 u_2 - u_{12}^2$. Here, $u_1$ and $u_2$ represent the quartic self-interaction coefficients of each field, while $u_{12}$ denotes the cross-interaction coupling. If $\Delta > 0$, the system exhibits a tetracritical point that allows for phase coexistence. Conversely, a value of $\Delta < 0$ drives the system toward a bicritical point, triggering an instability that strictly forbids coexistence enforcing a discontinuous, first-order phase transition \cite{altland2010condensed}.
\noindent In our case, when we factor out the coefficients of the quartic term from the functionals $\mathcal{F_R}$ and $\mathcal{F_H}$, the corresponding quantities are defined as follows:
\begin{equation}
    \begin{split}
        \Delta_R =4b_1b_2-b_3^2, \\ \Delta_H =4d_1d_2-d_3^2; 
    \end{split}
\label{discriminantes}
\end{equation}
\noindent by explicitly substituting the expressions for the fourth-order coefficients in terms of the thermal scale parameter given by \eqref{coefR} and \eqref{coefH} into the pair of equations \eqref{discriminantes}, we obtain the following equality:
\begin{equation}
    \Delta_R=\Delta_H=-\frac{\gamma^2(\gamma^2-1)^2}{36}(\lambda_R^2+\lambda_H^2).
\end{equation}
Since the $\lambda_i$-squared quantities are strictly positive, it follows directly that $\Delta_R=\Delta_H=\Delta < 0$, and consequently in this model, the strictly negative sign of $\Delta$ analytically confirms that the fields $\phi_1$ and $\phi_2$ cannot coexist, a restriction arising from a purely algebraic constraint. This result coupled with the exact equivalence of the effective masses at $T = T_c$, in which the pure circular symmetry $U(1)$ is recovered (where $\gamma = 0$), and formally establishes the existence of a bicritical point. 
\begin{figure}[H]
  \begin{center}
  \includegraphics[width=0.9\textwidth]{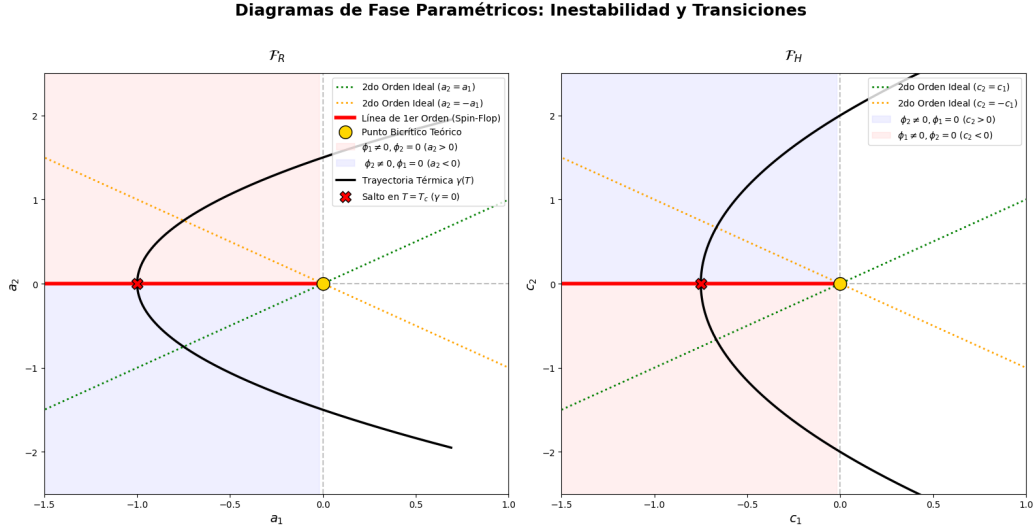}
\end{center}
\caption{Phase diagrams in the quadratic parameter space for the free energies functionals $\mathcal{F}_R$ (left) and the $\mathcal{F}_H$ (right). The shaded regions indicate the domains of the vacua ($\phi_1 \neq 0$ and $\phi_2 \neq 0$). The central red solid line marks the first-order phase transition boundary; the yellow dot denotes the theoretical bicritical point. The black solid curve represents the trajectory of $\gamma(T)$, while the red cross marks the transition at $T=T_c$.}
\label{fig5}
\end{figure}
\noindent Due to the configuration space of the free-energy density does not allow for a smooth, continuous deformation between preferred states, the vacuum transition—shifting from the stable minima on the $\phi_2$ axis to the $\phi_1$ axis upon crossing $T_c$—manifests inexorably as a discontinuous first-order phase transition. This mechanism is mathematically analogous to the metamagnetic reorientation observed in ``spin-flop'' phenomena \cite{fisher}.

\noindent The phase diagram in the effective mass space $(a_1, a_2)$ for $\mathcal{F}_R$ and equivalently $(c_1, c_2)$ for $\mathcal{F}_H$ is shown in Fig.\ref{fig5}. Because the geometric constraint on the hypercomplex invariants imposes a negative discriminant ($\Delta < 0$), the coexistence of mixed phases is strictly prohibited, which limits the boundary between the domains to a single line, formally establishing the existence of a bicritical point at the origin. Physically, the evolution of the system is confined to the parametric thermal trajectory $\gamma(T)$ (described by the black curve) imposed by the dissipative bath which intersects this boundary abruptly exactly at $T=T_c$.
This crossover represents a first-order discontinuous phase transition, it means a reorientation event similar to a spin-flop commented previously. In the real sector $\mathcal{F}_R$, the sign of the directional mass $a_2$ acts as the main breaking parameter: a value of $a_2 > 0$ stabilizes the oriented phase at $\phi_1$ (red region), while $a_2 < 0$ favors orientation along $\phi_2$ (blue region). Complementarily, the functional $\mathcal{F}_H$ exhibits an analogous geometry but with a parity inversion in its stability domains, demonstrating that the dissipative dynamics responds to the transition with an ordering in an opposite orientation, while preserving the structure of the bicritical singularity intact.
\section{Concluding remarks}
The construction of hypercomplex scalar electrodynamics under the local gauge symmetry $U(1) \times SO(1,1)$ establishes a rigorous theoretical framework where emerges as an intrinsic dynamic consequence of the ring algebra. To systematically investigate the phenomenological landscape of the associated phase transitions, a generalized Landau functional is constructed, which is constrained by the invariants permitted under this extended symmetry group. By applying the exact extremization criterion yields the generalized Ginzburg-Landau equations within the hypercomplex ring. These are extended into a system of four coupled differential equations, formulated explicitly in terms of the fundamental scalar field components and the electromagnetic potentials associated with both the subsystem and the environment. Physically, this generalized system describes the dynamics of the charged scalar fields which may correspond to charged bosonic particles and the exact macroscopic distribution of currents associated with the dissipative system.

In the subsequent analysis of the vacuum, grounded in the foundational principles of Landau's phenomenological theory, reveals important physical results. The algebraic parameter $\gamma$, which is initially introduced to reduce the degrees of freedom of the hypercomplex field to two real components, it is shown to inherently acquire the physical significance of a macroscopic thermal parameter. 
Through this formal identification, the thermodynamic analysis of dissipative scalar electrodynamics demonstrates the emergence of discontinuous first-order phase transitions and bicritical points that strictly preclude phase coexistence. Furthermore, the exact evaluation of the effective mass hierarchy, governed by the quadratic coefficients, details the behavior of the real $\mathcal{F}_R$ and hybrid $\mathcal{F}_H$ sector functionals; this analytical treatment formally proves that the orientation of the stable vacuum minima is dynamically selected across distinct temperature regimes.

As a future perspective, an immediate extension of this theoretical framework will focus on the formal incorporation of sixth-order invariants into the thermodynamic functional. This expansion will be instrumental in fully capturing the global stability of the vacuum and refining the exact analytical boundaries of these discontinuous phase transitions.

{\bf Acknowledgements:}
This work was supported by the Sistema Nacional de Investigadores (M\'exico). B.R. López Raymundo would like to thank Secretaría de Ciencia, Humanidades, Tecnología e Innovación (SECIHTI, México) for financial support (CVU: 2043835). The numerical analysis and graphics haven been made using Mathematica and codes developed in Python.

{\bf Data Availability Statement:} No data associated with the manuscript.

\end{document}